\documentclass[prb,twocolumn, superscriptaddress,floatfix]{revtex4}
\usepackage{graphicx}
\usepackage{epstopdf}
\usepackage{bm, amsmath, amssymb}
\usepackage{pdfpages}

\begin{document}

\title{Suppression of the radiative decay of atomic coherence in squeezed vacuum}

\author{K. W. Murch}
\affiliation{Quantum Nanoelectronics Laboratory, Department of Physics, University of California, Berkeley CA 94720}
\author{S. J. Weber}
\affiliation{Quantum Nanoelectronics Laboratory, Department of Physics, University of California, Berkeley CA 94720}
\author{K. M. Beck}
\affiliation{Department of Physics, MIT-Harvard Center for Ultracold Atoms, and Research Laboratory of Electronics, Massachusetts Institute of Technology, Cambridge, MA 02139}
\author{E. Ginossar}
\affiliation{Advanced Technology Institute and Department of Physics, University of Surrey, Guildford, GU2 7XH, United Kingdom}
\author{I. Siddiqi}
\affiliation{Quantum Nanoelectronics Laboratory, Department of Physics, University of California, Berkeley CA 94720}

\date{\today}

\maketitle

{\bf Quantum fluctuations of the electromagnetic vacuum are responsible for physical effects such as the Casimir force and the radiative decay of atoms, and set  fundamental limits on the sensitivity of measurements. Entanglement between photons can produce correlations that result in a reduction of these fluctuations  below the vacuum level allowing measurements that surpass the standard quantum limit in sensitivity.  \cite{trep02,gran87,xiao87,goda08,polz92}
 Here we demonstrate that the radiative decay rate of an atom that is coupled to  quadrature squeezed electromagnetic vacuum can be reduced below its natural linewidth.  We observe a two-fold reduction of the transverse radiative decay rate of a superconducting artificial atom coupled to continuum squeezed vacuum generated by a Josephson parametric amplifier, allowing the transverse coherence time $T_2$ to exceed the vacuum decay limit of $2T_1$.  We demonstrate that the measured radiative decay dynamics can be used to tomographically reconstruct the Wigner distribution of the the itinerant squeezed state.
Our results are the first confirmation of Gardiner's canonical prediction\cite{gard86} of quantum optics and open the door to new studies of the quantum light--matter interaction.}

\begin{figure}
\includegraphics[angle = 0, width = 0.5\textwidth]{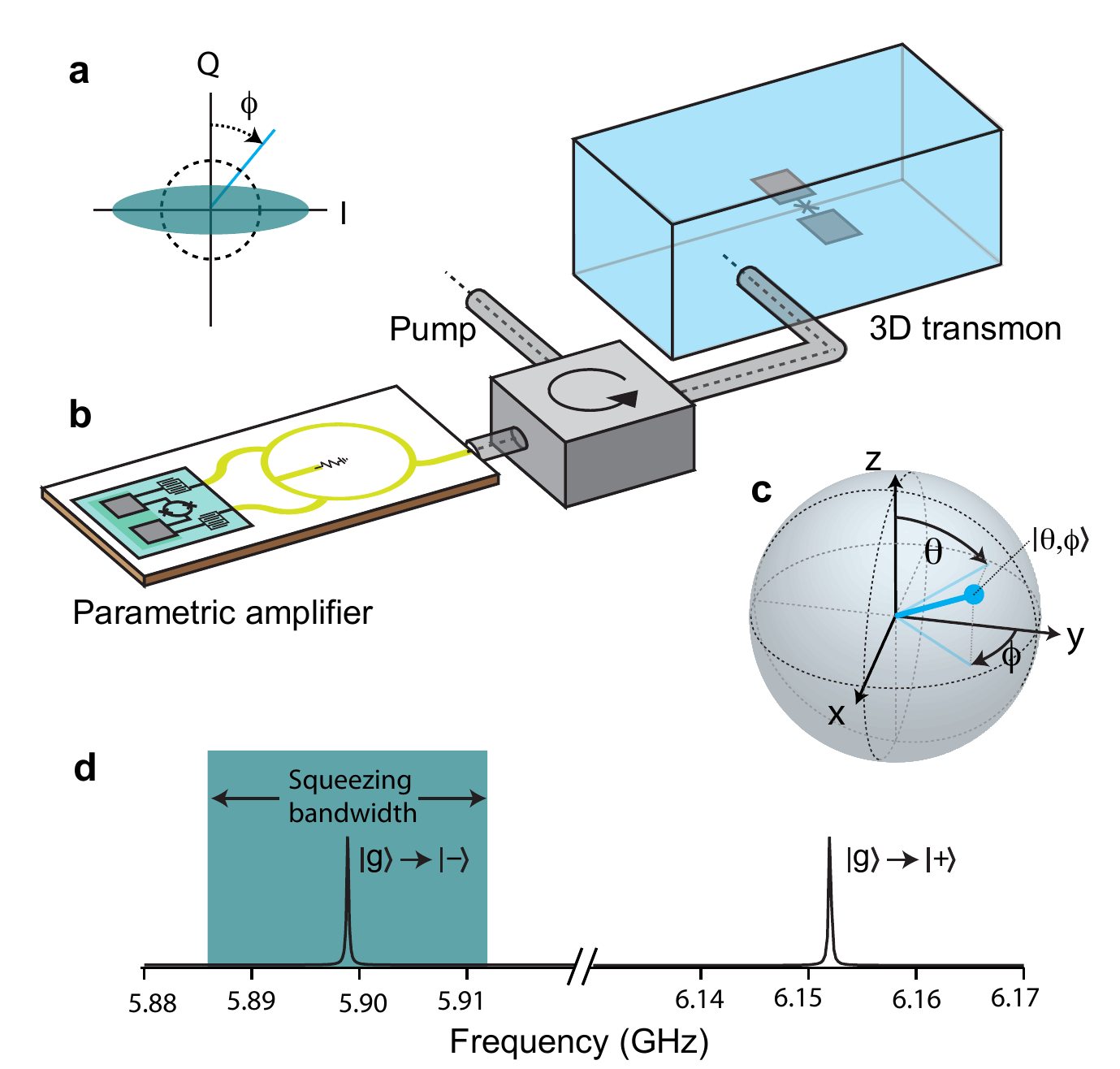}
\caption{\label{fig1} Experiment setup. {\bf a} The phase space of a mode of the electromagnetic field is described in terms of its in-phase ($I$) and quadrature phase ($Q$) components.  The Gaussian variance of the vacuum state is shown as a dashed line, and a squeezed state as the green region. {\bf b}  A lumped-element Josephson parametric amplifier is used to generate squeezed vacuum that is coupled to the input port of a 3D transmon qubit via a circulator with coaxial cables.  {\bf c} The state of a two level atom may be represented on the Bloch sphere with angles $\theta$ and $\phi$ describing the latitude and longitude respectively.   {\bf d} The resonant strong light-matter dipole interaction of the transmon circuit with the 3D cavity results in two polariton states $|+\rangle$ and $|-\rangle$.  The bandwidth of the squeezing is centered about the $|g\rangle \rightarrow |-\rangle$ transition frequency and is large compared to the natural linewidth of the transition.% The resonant Jaynes-Cummings interaction of the transmon circuit with the 3D cavity results in two polariton states $|+\rangle$ and $|-\rangle$.% The squeezed vacuum, $S(\omega)$, is resonant with the $|g\rangle \rightarrow |-\rangle$ transition.   
 }
\end{figure}

The quantization of the electromagnetic field implies a minimum uncertainty relation for non-commuting observables such as photon number and phase, or the in-phase ($I$) and quadrature phase ($Q$) amplitudes of a mode of the electromagnetic field.  The electromagnetic vacuum is a minimum uncertainty state with quantum fluctuations distributed equally between the two quadratures.  Parametric amplifiers operating in the optical \cite{slus85,ourj11,broo12} and microwave \cite{cast08,berg10,roch12,hatr11para,eich11} domain have been used to produce squeezed states of the electromagnetic field, wherein  the fluctuations in one quadrature  are increased and fluctuations in the canonically conjugate quadrature are reduced below the vacuum level, allowing for an improvement in measurement sensitivity. \cite{polz92,goda08,trep02,gran87,xiao87}  The focus of our research, however, is to reveal the effects of squeezed vacuum on the radiative properties of an atom.   In the optical domain, only a few experiments have explored the squeezed light--atom interaction, with studies in free space \cite{geor95,daya04} and in a cavity-QED architecture. \cite{turc98}  Our experiment takes place in the microwave domain and uses a polariton qubit---an effective two level atom  formed via the strong light-matter dipole interaction between a superconducting circuit and a microwave frequency cavity.  We use a Josephson parametric amplifier to produce  broadband squeezed vacuum in the modes of a transmission line that are resonant with the atomic transition. The architecture of a one-dimensional radiative environment\cite{gard86,park93,turc98,gino05} and the strong coupling available in circuit-QED\cite{scho10} enable us to engineer the radiative decay to be solely into the modes of the transmission line that are occupied by squeezed vacuum.  Thus we are able to systematically explore the dynamics of the atom under the modified radiative reservoir.

The itinerant electromagnetic field generated by a degenerate parametric amplifier may be approximately described in terms of the squeezing moments, $M$ and $N$, that are related to the frequency correlations of the output field. In the limit of large amplifier bandwidth $\langle a^\dagger(\omega) a(\omega')\rangle  = N \delta(\omega-\omega')$ and $ \langle a(\omega) a(\omega')\rangle  = M \delta(\omega + \omega' - 2 \omega_0)$, where $a,a^\dagger$ are the creation and annihilation operators of the output field of the amplifier, and $\omega_0$ is the center frequency of the amplifier.   Squeezed states occur when  $M>N$, with $M$ bounded by $M^2 \leq N(N+1)$. The radiative decay dynamics of an atom that couples to a broadband squeezed reservoir centered about the atomic transition frequency are governed by the optical Gardiner-Bloch equations,\cite{gard86} %I think this name already appears in literature 
\begin{eqnarray}
\langle \dot{\sigma}_x \rangle  = -\gamma (N-M+1/2) \langle \sigma_x\rangle, \nonumber\\
\langle \dot{\sigma}_y \rangle  = -\gamma (N+M+1/2) \langle \sigma_y\rangle ,\label{eq:obe}\\
\langle \dot{\sigma}_z \rangle  = -\gamma (2 N+1) \langle \sigma_z\rangle + \gamma  \nonumber.
\end{eqnarray}
Here, $\sigma_x,\ \sigma_y,$ and, $\sigma_z$ are the pseudospin operators for a two-level atom.  As shown in Figure 1, the $Q$ quadrature of the electromagnetic vacuum is squeezed, and a coherent drive along this axis induces rotations of the atom about the $y$ axis of the Bloch sphere.
By setting $M,N=0$ in Eq. 1 we recover the case of radiative decay into electromagnetic vacuum, where the transverse coherence decays half as fast as the longitudinal coherence ($T_2 = 2T_1 = 2/\gamma$).   In contrast, the radiative decay into squeezed vacuum is characterized by timescales, $\tilde{T}_x =T_1/(N-M+1/2),\ \tilde{T}_y = T_1/(N+M+1/2)$, and $T_z = T_1/(2N+1)$.   Specifically, in the limit of large squeezing it is predicted that the transverse decay time $\tilde{T}_x$ is increased beyond the value of $2T_1$ owing to the reduced fluctuations in the $Q$ quadrature of the vacuum.

   \begin{figure}
\includegraphics[angle = 0, width = .5\textwidth]{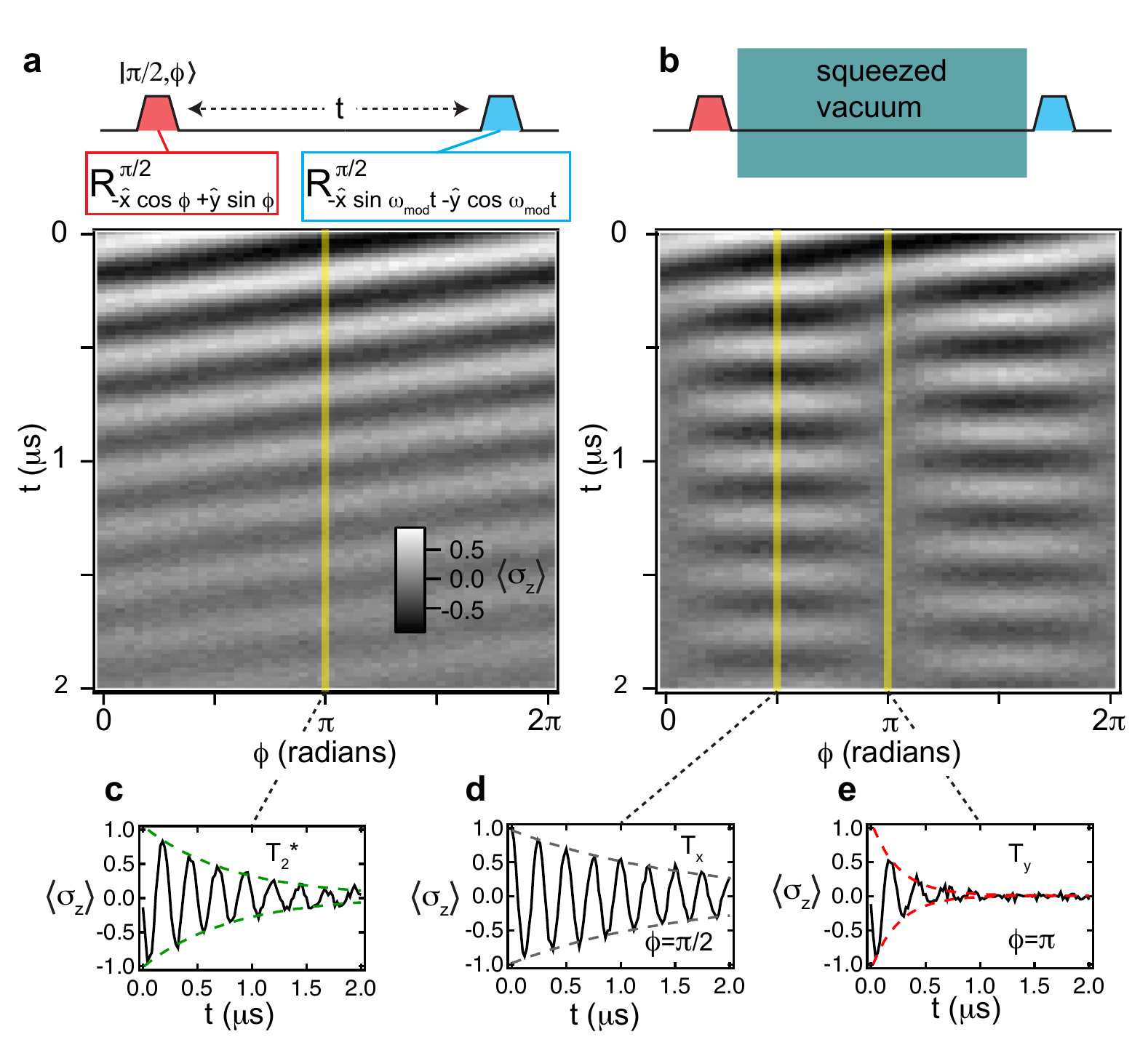}
\caption{\label{fig2} Transverse decay into squeezed vacuum. {\bf a} The Ramsey measurement as a function of  angle consisted of a first $\pi/2$ rotation about  the $-\hat{x}\cos \phi +\hat{y} \sin \phi$ axis, to prepare the in the state $|\pi/2,\phi\rangle$,  followed by a second $\pi/2$ rotation about the $-\hat{x}\sin \omega_\textrm{mod} t-\hat{y} \cos \omega_{\textrm{mod}} t$ axis applied at variable time $t$. The 2D plot displays $\langle \sigma_z\rangle$ as a function of $t$ and $\phi$ which is characterized by sinusoidal decay with a uniform decay constant $T_2^*$ and phase $\phi$. {\bf b} The transverse decay into squeezed vacuum was measured by turning the pump for the LJPA on  interim the qubit pulses.  The 2D plot indicates that after rapid decay of coherence along the $\pm\hat{y}$ axes, the resulting coherence along the $\pm\hat{x}$ axes decays with time constant $T_x>T_2^*$. {\bf c} The  Ramsey measurement for the qubit prepared along the  $-\hat{y}$ ($\phi = \pi$) axis with the squeezing off. {\bf d,e} The Ramsey measurement in the presence of squeezed vacuum for the qubit prepared along the  $-\hat{y}$ ($\phi = \pi$), and  $+\hat{x}$ ($\phi = \pi/2$) axes. }
\end{figure}

A simplified schematic of our experiment is shown in Figure 1b.  We realized an effective two level system using the ground state and lower energy level of a polariton formed by a superconducting transmon\cite{koch07transmon} circuit resonantly coupled to the TE$_{101}$ mode of a 3D superconducting cavity.\cite{paik113D}    The transition frequency of the effective qubit  was $\omega_\mathrm{q}/2\pi = 5.8989$ GHz with $T_1 = 0.65(2)$ $\mu$s set by deliberate coupling to the 50 $\Omega$ environment. In the supplementary information we show in detail that the radiative interaction of the polariton with squeezed vacuum is that of an idealized atom interacting directly with a squeezed reservoir.  Squeezed vacuum was generated by pumping a lumped-element Josephson Parametric amplifier (LJPA) with two tones at frequencies $\omega_1$ and $\omega_2$ that were evenly spaced about the qubit transition frequency\cite{kama09}, and satisfied $\omega_{0} = (\omega_1+\omega_2)/2 = \omega_\mathrm{q}$.  The bandwidth of the squeezing was 13 MHz, sufficient to fulfill the large bandwidth assumption based on the radiative linewidth of the qubit $\gamma/2\pi = 240$ kHz. 
 The output of the amplifier was connected with coaxial cables to the strongly coupled port of the superconducting cavity.  %Due to the finite temperature of the 50 $\Omega$ environment and other sources of noise, a small average number of photons $N_\textrm{th} \leq 0.019$ (see methods)contaminated the vacuum environment of the qubit.   we measured an average $N_\textrm{th} \leq 0.019$ photons that contaminate the vacuum environment of the qubit.  While small, these thermal photons degrade the squeezing and are included in our analysis.  %Due to the finite temperature of the 50 $\Omega$ environment and other sources of noise, we expect a small average number of photons per mode, $N_\textrm{th}$, to contaminate the vacuum environment of the qubit.  This bath of  thermal photons both reduces the measured energy decay time $T_1$ from its intrinsic value by a factor of $1/(2N_\textrm{th}+1)$, and increases the equilibrium excited state population.  We determined the equilibrium excited state population to be $1.8\%$ using a Rabi population measurement\cite{geer12}, allowing us to place a pessimistic limit on the number of thermal photons that characterize our vacuum environment of $N_\textrm{th} \leq 0.019$ and thus the intrinsic radiative decay time $T_1\leq0.67 \ \mu$s.

To demonstrate the effect of squeezed vacuum on the transverse decay of the qubit we conducted Ramsey measurements at different angles along the equator of the Bloch sphere.   The Ramsey measurements consisted of an initial $\pi/2$ rotation about  the $-\hat{x}\cos \phi +\hat{y} \sin \phi$ axis, followed by a second $\pi/2$ rotation about the $-\hat{x}\sin \omega_\textrm{mod} t-\hat{y} \cos \omega_{\textrm{mod}} t$ axis applied at variable time $t$. Modulation of the rotation angle of the second $\pi/2$ pulse at frequency $\omega_{\textrm{mod}}$ results in oscillatory Ramsey fringes without detuning.  Figure 2a displays $\langle \sigma_z\rangle$ as a function of time and angle with the squeezing turned off; $\langle \sigma_z\rangle$ exhibits exponentially damped, sinusoidal oscillations at angular frequency $\omega_{\textrm{mod}}$ and phase $\phi$, with a uniform decay time $T_2^* = 1.08(4)\ \mu$s. The reduction of the $T_2^*$ time from $2T_1$ indicates the presence of a small amount of pure dephasing characterized by a time scale $T_\phi = 6.6(5) \ \mu$s.
Figure 2b displays the results of the Ramsey measurement when the LJPA pump was turned on to generate squeezed vacuum for the variable duration between the first and second $\pi/2$ pulses. The power gain of the amplifier was 4 dB.  The transverse decay in the presence of squeezed vacuum reveals two timescales, $T_x = 1.67 \ \mu$s, and $T_y =0.28\ \mu$s, that describe the exponential decay of coherence when the qubit was prepared along the $\pm \hat{x}$ and $\pm \hat{y}$ axes respectively. Subtracting the pure dephasing from the measured timescales gives the radiative transverse decay times, $\tilde{T}_x = 2.2 \ \mu$s, and $\tilde{T}_y = 0.29\ \mu$s. The interaction with squeezed vacuum both enhances decay along the $\hat{y}$ axis due to the increased fluctuations in the $I$ quadrature of the field, and suppresses decay along the $\hat{x}$ axis due to the reduced fluctuations in $Q$.

   \begin{figure*}[htb]
\includegraphics[angle = 0, width = .8\textwidth]{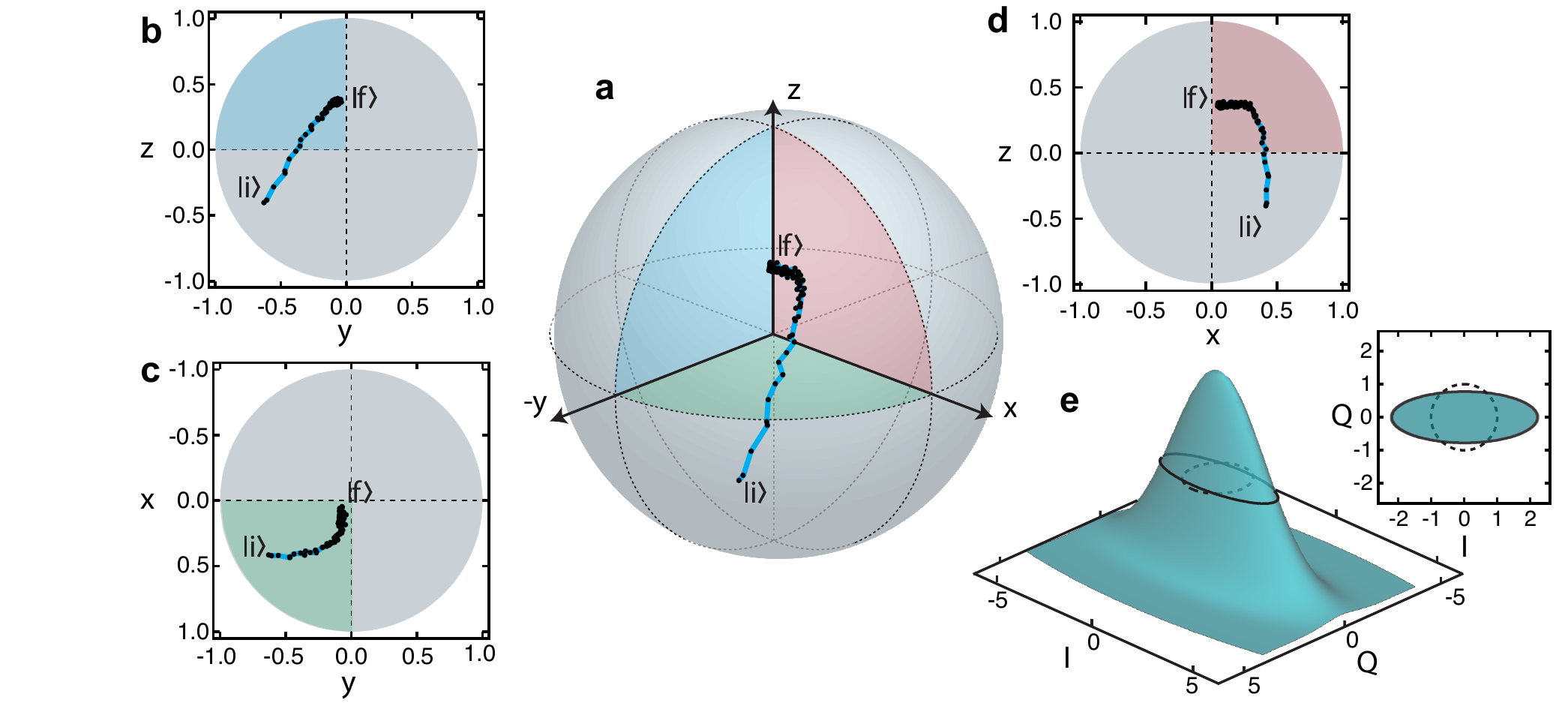}
\caption{Radiative decay dynamics in squeezed vacuum.  {\bf a-d} Quantum state tomography shows the evolution of the Bloch vector which was initialized at $|i\rangle = |0.67\pi,0.83\pi\rangle$ with tomographic measurements equally spaced between 0 and 3 $\mu$s.  The dynamics are characterized by fast decay along $\hat{y}$ and $\hat{z}$ with slow decay along $\hat{x}$. {\bf e} From the radiative decay rates we tomographically reconstruct the Wigner quasiprobability distribution of the itinerant squeezed vacuum mode at $\omega_0$. The inset shows the  Gaussian half width of the squeezed  and vacuum states as solid and dashed lines respectively.\label{fig:tomo}}
\end{figure*}

   \begin{figure*}
\includegraphics[angle = 0, width = .8\textwidth]{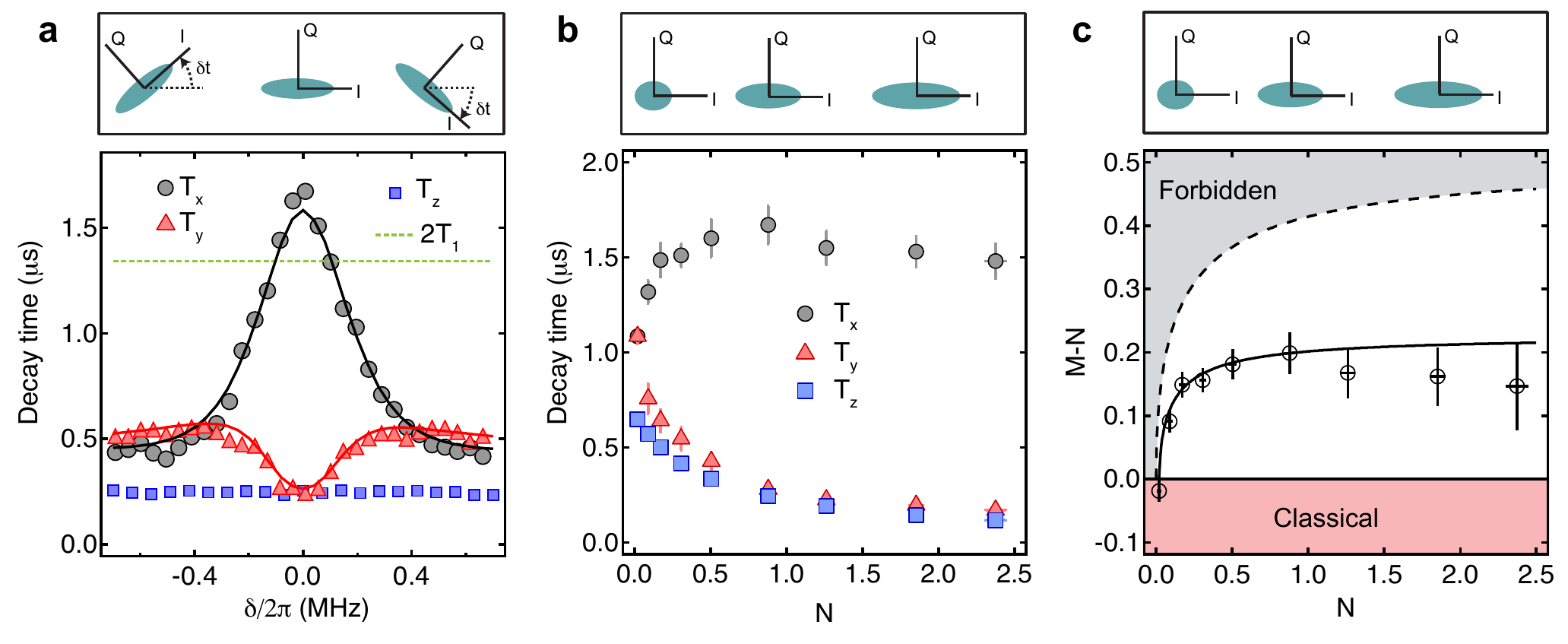}
\caption{Dependence of the transverse and longitudinal decay times on LJPA detuning and bias.   {\bf a} Effective decay constants vs the detuning of the center frequency of the LJPA from the qubit, $\delta = \omega_0 - \omega_\mathrm{q}$. $T_x$ and $T_y$ show a dependence on the detuning of the squeezing from the qubit transition frequency, reaching their maximum and minimum values on resonance.    The vacuum decay limit, $2T_1$  is shown for comparison. The solid black and red lines are the theoretical dependence of $T_x$ and $T_y$ on the detuning. The upper cartoon indicates how the detuning of the squeezing causes the squeezing ellipse to rotate relative to the qubit coordinates.  {\bf b} Measured values for $T_x,\ T_y, \ T_z$ for increasing gain of the LJPA vs $N$. Error bars indicate the standard error of the mean based on 10 successive measurements.  The upper inset indicates how the aspect ratio of the squeezed state changes for increasing $N$.  {\bf c} $M-N$ is plotted versus $N$.  The dashed line indicates a minimum uncertainty squeezed state that is  expected  for ideal squeezing.  The solid line indicates the expected dependence for a quantum efficiency of $\eta = 0.5$. The gray region indicates values of $M$ and $N$ that are forbidden and the red region indicate values of $M$ and $N$ that correspond to classical states of light. \label{fig:freq} }
\end{figure*}

The radiative decay dynamics in the presence of squeezed vacuum can be presented as a trajectory of the Bloch vector.  To illustrate this point, we prepared the qubit in an initial state $|i\rangle$ with a $0.67 \pi$ rotation about the -$\hat{x}\sin \phi +\hat{y} \cos \phi$ axis with $\phi = 0.32 \pi$. After this preparation, the pump of the LJPA was turned on for a variable period of time to generate squeezed vacuum.  After this variable duration, we tomographically reconstructed the qubit state using $\pi/2$ rotations around the $\hat{x}$ and $\hat{y}$ axes followed by state readout in the $\sigma_z$ basis to determine the Bloch vector components $\langle \sigma_y\rangle$ and $\langle \sigma_x\rangle$, or no rotation to determine $\langle \sigma_z\rangle$.      The trajectory of the Bloch vector, displayed in Figure \ref{fig:tomo}, follows the expected decay dynamics based on (\ref{eq:obe}) with fast decay along the $\hat{y}$ and $\hat{z}$ axes and by slow relaxation along $\hat{x}$.  The final state of the qubit is described by $\langle \sigma_x \rangle \rightarrow 0$, $\langle \sigma_y\rangle \rightarrow 0.07$, $\langle \sigma_z\rangle \rightarrow 0.36$.  The steady state value of $\langle \sigma_z\rangle$ is consistent with a bath of  $N = 0.88$ photons that characterize average photon occupation of the squeezed state.  The remnant coherence along the $\hat{y}$ axis is the result of a small coherent component of the squeezed state.  This coherent drive, characterized by a Rabi frequency $\Omega_\mathrm{R}\ll1/T_1$, in combination with the radiative decay of the qubit results in a steady state coherence, \cite{carm87} $\langle \sigma_y\rangle_\mathrm{ss} \propto \Omega_\textrm{R} T_1$.   Based on our measurements,  $\Omega_\mathrm{R}\simeq 2\pi \times 10$ kHz, consistent with a 65 dB on/off ratio of our qubit manipulation pulses.

Because the qubit's decay dynamics are sensitive to altered vacuum fluctuations, they can be used to probe squeezed states of light.  Previously, noise and correlation  measurements have been used to characterize the squeezed states generated by microwave parametric amplifiers. \cite{mall11,eich11, berg12,flur12} Similarly, qubits have been used to tomographically reconstruct localized nonclassical states of light. \cite{hofh09,kirc12}   Here we use the qubit's decay dynamics to tomographically reconstruct, to second order, the Wigner distribution for the itinerant squeezed state generated by the LJPA.  From  $T_z$, the measured decay constant of $\langle \sigma_z\rangle$ and $\tilde{T_x}$ we determine $N=0.88$ and $M=1.08$, from which the Gaussian variances  $\sigma_I^2 = 2(N+M+1/2)$ and $\sigma_Q^2 = 2(N-M+1/2)$ are calculated.  Figure \ref{fig:tomo}e displays the reconstructed Wigner distribution for the squeezed mode at frequency $\omega_0$.

  In Figure \ref{fig:freq}a, we display the effective decay constants for different values of the detuning $\delta =\omega_0 -  \omega_\mathrm{q}$ between the center frequency of the pump and the qubit.  $T_x$ and $T_y$ depend strongly on the detuning near resonance highlighting the ``smoking gun" evidence of interaction with squeezed vacuum where $T_x>2T_1$.  When the detuning is large, the squeezing axis rotates rapidly with respect to the qubit axis and the decay times approach a constant value of $2T_1/(2N+1)$.  The solid black and red lines indicate the expected dependence of $T_x$ and $T_y$ on $\delta$ as discussed in the supplementary information.

  In Figure \ref{fig:freq}b we display the decay constants measured for different bias conditions of the LJPA obtained by changing the power of the pump tones.  The transverse decay rates were measured as depicted in Figure \ref{fig2}.   As expected,  larger gain of the LJPA results in larger amounts of squeezing with an associated increase of $T_x$ and decrease  of $T_y$ and $T_z$.    Figure \ref{fig:freq}c displays $M-N$ versus $N$.   The reduction of $M$ from its maximum allowed value, shown as a dashed line, may be attributed to two possible sources; losses in the microwave components between the LJPA and the qubit, and non-ideal performance of the LJPA.  If we assume that the LJPA produces an ideal squeezed state, with $M = \sqrt{N(N+1)}$, then the degradation can be accounted for by an attenuation of the squeezed vacuum from the LJPA by a factor of $\eta = 0.5$.  Attenuation degrades the squeezed vacuum by absorbing correlated photons thereby making the quadrature  fluctuations tend toward the normal vacuum fluctuations.  This level of attenuation is consistent with the anticipated insertion loss between the LJPA and the qubit due to the microwave components we used.  At values of $N>1$, however, it appears that the performance of the LJPA may become non-ideal as indicated by the slight reduction of $M-N$ for $N>1$.

Our results demonstrate the ability to alter the vacuum environment of a two level atom to a degree that has so far been elusive in atomic and molecular systems, allowing the direct study of a long sought physical phenomenon of the light-matter interaction.     Our system also demonstrates the strength of using superconducting artificial atoms as sensitive detectors of the quantum states of the electromagnetic field.  Future studies with squeezed light and superconducting qubits may enhance the fidelity of quantum gates, enable the generation of multi-qubit entanglement \cite{hald99} and enable the study of non-Markovian quantum baths.

\small {\bf Methods Summary}  

Due to the finite temperature of the 50 $\Omega$ environment and other sources of noise, a small average number of photons, $N_\textrm{th}$, are expected to contaminate the vacuum environment of the qubit.  This bath of  thermal photons both reduces the measured energy decay time $T_1$ from its intrinsic value by a factor of $1/(2N_\textrm{th}+1)$, and increases the equilibrium excited state population.  We determined the equilibrium excited state population to be $1.8\%$ using a Rabi population measurement\cite{geer12}, allowing us to place a pessimistic limit on the number of thermal photons that characterize our vacuum environment of $N_\textrm{th} \leq 0.019$ and thus the intrinsic radiative decay time $T_1\leq0.67 \ \mu$s.  While a small effect, these thermal photons were included in our determination of $N$ and $M$.

The polariton qubit was composed of a transmon circuit with charging energy $E_\mathrm{C}/h=208$ MHz and Josephson energy $E_\mathrm{J}/h = 23.27$ GHz coupled to a 3D aluminum cavity with resonance frequency $\omega_\mathrm{c}/2\pi = 6.0456$ GHz at rate $g/2\pi = 126$ MHz.  The cavity was equipped with two ports; one strongly coupled port that limited the quality factor to  $Q =1.1\times10^4$ and another weakly coupled port. The qubit was enclosed in successive layers of superconducting and magnetic shields and  anchored to the mixing chamber stage of a dilution refrigerator with a base temperature of $20$ mK.  State readout was performed using the Jaynes-Cummings nonlinearity technique\cite{reed10} by driving the weakly coupled port of the cavity with a strong tone at 6.0467 GHz and integrating the first 200 ns of transmitted  signal.  

The LJPA was composed of a two junction SQUID formed of $1\ \mu$A Josephson junctions shunted with $1$ pF of capacitance and isolated from the input ports of a 180$^\circ$ hybrid with interdigitated capacitors that resulted in a quality factor of $Q_\mathrm{LJPA} = 100$.  The LJPA was flux biased to have a low power resonance at $5.897$ GHz. The differential port of the hybrid was connected to the strongly coupled port of the qubit with coaxial lines via two circulators and a -20 dB coupler that allowed the injection of the qubit manipulation pulses.

A single microwave source was used to generate the qubit preparation, tomography, and LJPA pump pulses.  The LJPA pump was obtained by driving an $IQ$ mixer with a tone at 540 MHz, and adjusting the dc offsets to null the carrier.

{\bf Acknowledgments}

We thank C. Macklin, N. Roch, and Lev S. Bishop for useful discussions.  This research was supported in part (K.M, S.W., and I.S.) by the Office of Naval Research (ONR) and the Office of the Director of National Intelligence (ODNI), Intelligence Advanced Research Projects Activity (IARPA), through the Army Research Office. All statements of fact, opinion or conclusions contained herein are those of the authors and should not be construed as representing the official views or policies of IARPA, the ODNI, or the US Government. E.G. acknowledges support from EPSRC (EP/I026231/1). K.B.  acknowledges support from NSF GRFP (0645960) and IGERT (0801525).

%\bibliographystyle{prsty}
%\bibliographystyle{naturemag}
%\bibliography{squeezingrefs}

\begin{thebibliography}{10}
\expandafter\ifx\csname url\endcsname\relax
  \def\url#1{\texttt{#1}}\fi
\expandafter\ifx\csname urlprefix\endcsname\relax\def\urlprefix{URL }\fi
\providecommand{\bibinfo}[2]{#2}
\providecommand{\eprint}[2][]{\url{#2}}

\bibitem{trep02}
\bibinfo{author}{Treps, N.} \emph{et~al.}
\newblock \bibinfo{title}{Surpassing the standard quantum limit for optical
  imaging using nonclassical multimode light}.
\newblock \emph{\bibinfo{journal}{Phys. Rev. Lett.}}
  \textbf{\bibinfo{volume}{88}}, \bibinfo{pages}{203601}
  (\bibinfo{year}{2002}).

\bibitem{gran87}
\bibinfo{author}{Grangier, P.}, \bibinfo{author}{Slusher, R.~E.},
  \bibinfo{author}{Yurke, B.} \& \bibinfo{author}{LaPorta, A.}
\newblock \bibinfo{title}{Squeezed-light enhanced polarization interferometer}.
\newblock \emph{\bibinfo{journal}{Phys. Rev. Lett.}}
  \textbf{\bibinfo{volume}{59}}, \bibinfo{pages}{2153--2156}
  (\bibinfo{year}{1987}).

\bibitem{xiao87}
\bibinfo{author}{Xiao, M.}, \bibinfo{author}{Wu, L.-A.} \&
  \bibinfo{author}{Kimble, H.~J.}
\newblock \bibinfo{title}{Precision measurement beyond the shot-noise limit}.
\newblock \emph{\bibinfo{journal}{Phys. Rev. Lett.}}
  \textbf{\bibinfo{volume}{59}}, \bibinfo{pages}{278--281}
  (\bibinfo{year}{1987}).

\bibitem{goda08}
\bibinfo{author}{Goda, K.} \emph{et~al.}
\newblock \bibinfo{title}{A quantum-enhanced prototype gravitational-wave
  detector}.
\newblock \emph{\bibinfo{journal}{Nature Physics}}
  \textbf{\bibinfo{volume}{4}}.

\bibitem{polz92}
\bibinfo{author}{Polzik, E.~S.}, \bibinfo{author}{Carri, J.} \&
  \bibinfo{author}{Kimble, H.~J.}
\newblock \bibinfo{title}{Spectroscopy with squeezed light}.
\newblock \emph{\bibinfo{journal}{Phys. Rev. Lett.}}
  \textbf{\bibinfo{volume}{68}}, \bibinfo{pages}{3020--3023}
  (\bibinfo{year}{1992}).

\bibitem{gard86}
\bibinfo{author}{Gardiner, C.~W.}
\newblock \bibinfo{title}{Inhibition of atomic phase decays by squeezed light:
  A direct effect of squeezing}.
\newblock \emph{\bibinfo{journal}{Phys. Rev. Lett.}}
  \textbf{\bibinfo{volume}{56}}, \bibinfo{pages}{1917--1920}
  (\bibinfo{year}{1986}).

\bibitem{slus85}
\bibinfo{author}{Slusher, R.~E.}, \bibinfo{author}{Hollberg, L.~W.},
  \bibinfo{author}{Yurke, B.}, \bibinfo{author}{Mertz, J.~C.} \&
  \bibinfo{author}{Valley, J.~F.}
\newblock \bibinfo{title}{Observation of squeezed states generated by four-wave
  mixing in an optical cavity}.
\newblock \emph{\bibinfo{journal}{Phys. Rev. Lett.}}
  \textbf{\bibinfo{volume}{55}}, \bibinfo{pages}{2409--2412}
  (\bibinfo{year}{1985}).

\bibitem{ourj11}
\bibinfo{author}{Ourjoumtsev, A.} \emph{et~al.}
\newblock \bibinfo{title}{Observation of squeezed light from one atom excited
  with two photons}.
\newblock \emph{\bibinfo{journal}{Nature}} \textbf{\bibinfo{volume}{474}},
  \bibinfo{pages}{623--626} (\bibinfo{year}{2011}).

\bibitem{broo12}
\bibinfo{author}{Brooks, D.} \emph{et~al.}
\newblock \bibinfo{title}{Non-classical light generated by quantum-noise-driven
  cavity optomechanics}.
\newblock \emph{\bibinfo{journal}{Nature}} \textbf{\bibinfo{volume}{488}},
  \bibinfo{pages}{476--480} (\bibinfo{year}{2012}).

\bibitem{cast08}
\bibinfo{author}{Castellanos-Beltran, M.~A.}, \bibinfo{author}{Irwin, K.~D.},
  \bibinfo{author}{Hilton, G.~C.}, \bibinfo{author}{Vale, L.~R.} \&
  \bibinfo{author}{Lehnert, K.~W.}
\newblock \bibinfo{title}{Amplification and squeezing of quantum noise with a
  tunable josephson metamaterial}.
\newblock \emph{\bibinfo{journal}{Nature Physics}}
  \textbf{\bibinfo{volume}{4}}, \bibinfo{pages}{929--931}
  (\bibinfo{year}{2008}).

\bibitem{berg10}
\bibinfo{author}{Bergeal, N.} \emph{et~al.}
\newblock \bibinfo{title}{Phase preserving amplification near the quantum limit
  with a josephson ring modulator}.
\newblock \emph{\bibinfo{journal}{Nature}} \textbf{\bibinfo{volume}{465}},
  \bibinfo{pages}{64--68} (\bibinfo{year}{2010}).

\bibitem{roch12}
\bibinfo{author}{Roch, N.} \emph{et~al.}
\newblock \bibinfo{title}{Widely tunable, nondegenerate three-wave mixing
  microwave device operating near the quantum limit}.
\newblock \emph{\bibinfo{journal}{Phys. Rev. Lett.}}
  \textbf{\bibinfo{volume}{108}}, \bibinfo{pages}{147701}
  (\bibinfo{year}{2012}).

\bibitem{hatr11para}
\bibinfo{author}{Hatridge, M.}, \bibinfo{author}{Vijay, R.},
  \bibinfo{author}{Slichter, D.~H.}, \bibinfo{author}{Clarke, J.} \&
  \bibinfo{author}{Siddiqi, I.}
\newblock \bibinfo{title}{Dispersive magnetometry with a quantum limited squid
  parametric amplifier}.
\newblock \emph{\bibinfo{journal}{Phys. Rev. B}} \textbf{\bibinfo{volume}{83}},
  \bibinfo{pages}{134501} (\bibinfo{year}{2011}).

\bibitem{eich11}
\bibinfo{author}{Eichler, C.} \emph{et~al.}
\newblock \bibinfo{title}{Observation of two-mode squeezing in the microwave
  frequency domain}.
\newblock \emph{\bibinfo{journal}{Phys. Rev. Lett.}}
  \textbf{\bibinfo{volume}{107}}, \bibinfo{pages}{113601}
  (\bibinfo{year}{2011}).

\bibitem{geor95}
\bibinfo{author}{Georgiades, N.~P.}, \bibinfo{author}{Polzik, E.~S.},
  \bibinfo{author}{Edamatsu, K.}, \bibinfo{author}{Kimble, H.~J.} \&
  \bibinfo{author}{Parkins, A.~S.}
\newblock \bibinfo{title}{Nonclassical excitation for atoms in a squeezed
  vacuum}.
\newblock \emph{\bibinfo{journal}{Phys. Rev. Lett.}}
  \textbf{\bibinfo{volume}{75}}, \bibinfo{pages}{3426--3429}
  (\bibinfo{year}{1995}).

\bibitem{daya04}
\bibinfo{author}{Dayan, B.}, \bibinfo{author}{Pe'er, A.},
  \bibinfo{author}{Friesem, A.~A.} \& \bibinfo{author}{Silberberg, Y.}
\newblock \bibinfo{title}{Two photon absorption and coherent control with
  broadband down-converted light}.
\newblock \emph{\bibinfo{journal}{Phys. Rev. Lett.}}
  \textbf{\bibinfo{volume}{93}}, \bibinfo{pages}{023005}
  (\bibinfo{year}{2004}).

\bibitem{turc98}
\bibinfo{author}{Turchette, Q.~A.}, \bibinfo{author}{Georgiades, N.~P.},
  \bibinfo{author}{Hood, C.~J.}, \bibinfo{author}{Kimble, H.~J.} \&
  \bibinfo{author}{Parkins, A.~S.}
\newblock \bibinfo{title}{Squeezed excitation in cavity qed:\quad{}experiment
  and theory}.
\newblock \emph{\bibinfo{journal}{Phys. Rev. A}} \textbf{\bibinfo{volume}{58}},
  \bibinfo{pages}{4056--4077} (\bibinfo{year}{1998}).

\bibitem{park93}
\bibinfo{author}{Parkins, A.~S.}, \bibinfo{author}{Zoller, P.} \&
  \bibinfo{author}{Carmichael, H.~J.}
\newblock \bibinfo{title}{Spectral linewidth narrowing in a strongly coupled
  atom-cavity system via squeezed-light excitation of a ``vacuum'' rabi
  resonance}.
\newblock \emph{\bibinfo{journal}{Phys. Rev. A}} \textbf{\bibinfo{volume}{48}},
  \bibinfo{pages}{758--763} (\bibinfo{year}{1993}).

\bibitem{gino05}
\bibinfo{author}{Ginossar, E.} \& \bibinfo{author}{Levit, S.}
\newblock \bibinfo{title}{Semiconductor microstructure in a squeezed vacuum:
  Electron-hole plasma luminescence}.
\newblock \emph{\bibinfo{journal}{Phys. Rev. B}} \textbf{\bibinfo{volume}{72}},
  \bibinfo{pages}{075333} (\bibinfo{year}{2005}).

\bibitem{scho10}
\bibinfo{author}{Girvin, R. J. S. . S.~M.}
\newblock \bibinfo{title}{Wiring up quantum systems}.
\newblock \emph{\bibinfo{journal}{Nature}} \textbf{\bibinfo{volume}{451}},
  \bibinfo{pages}{664--669} (\bibinfo{year}{2008}).

\bibitem{koch07transmon}
\bibinfo{author}{Koch, J.} \emph{et~al.}
\newblock \bibinfo{title}{Charge-insensitive qubit design derived from the
  cooper pair box}.
\newblock \emph{\bibinfo{journal}{Phys. Rev. A}} \textbf{\bibinfo{volume}{76}},
  \bibinfo{pages}{042319} (\bibinfo{year}{2007}).

\bibitem{paik113D}
\bibinfo{author}{Paik, H.} \emph{et~al.}
\newblock \bibinfo{title}{Observation of high coherence in josephson junction
  qubits measured in a three-dimensional circuit qed architecture}.
\newblock \emph{\bibinfo{journal}{Phys. Rev. Lett.}}
  \textbf{\bibinfo{volume}{107}}, \bibinfo{pages}{240501}
  (\bibinfo{year}{2011}).

\bibitem{kama09}
\bibinfo{author}{Kamal, A.}, \bibinfo{author}{Marblestone, A.} \&
  \bibinfo{author}{Devoret, M.}
\newblock \bibinfo{title}{Signal-to-pump back action and self-oscillation in
  double-pump josephson parametric amplifier}.
\newblock \emph{\bibinfo{journal}{Phys. Rev. B}} \textbf{\bibinfo{volume}{79}},
  \bibinfo{pages}{184301} (\bibinfo{year}{2009}).

\bibitem{carm87}
\bibinfo{author}{Carmichael, H.~J.}, \bibinfo{author}{Lane, A.~S.} \&
  \bibinfo{author}{Walls, D.~F.}
\newblock \bibinfo{title}{Resonance fluorescence from an atom in a squeezed
  vacuum}.
\newblock \emph{\bibinfo{journal}{Phys. Rev. Lett.}}
  \textbf{\bibinfo{volume}{58}}, \bibinfo{pages}{2539--2542}
  (\bibinfo{year}{1987}).

\bibitem{mall11}
\bibinfo{author}{Mallet, F.} \emph{et~al.}
\newblock \bibinfo{title}{Quantum state tomography of an itinerant squeezed
  microwave field}.
\newblock \emph{\bibinfo{journal}{Phys. Rev. Lett.}}
  \textbf{\bibinfo{volume}{106}}, \bibinfo{pages}{220502}
  (\bibinfo{year}{2011}).

\bibitem{berg12}
\bibinfo{author}{Bergeal, N.}, \bibinfo{author}{Schackert, F.},
  \bibinfo{author}{Frunzio, L.} \& \bibinfo{author}{Devoret, M.~H.}
\newblock \bibinfo{title}{Two-mode correlation of microwave quantum noise
  generated by parametric down-conversion}.
\newblock \emph{\bibinfo{journal}{Phys. Rev. Lett.}}
  \textbf{\bibinfo{volume}{108}}, \bibinfo{pages}{123902}
  (\bibinfo{year}{2012}).

\bibitem{flur12}
\bibinfo{author}{Flurin, E.}, \bibinfo{author}{Roch, N.},
  \bibinfo{author}{Mallet, F.}, \bibinfo{author}{Devoret, M.~H.} \&
  \bibinfo{author}{Huard, B.}
\newblock \bibinfo{title}{Generating entangled microwave radiation over two
  transmission lines}.
\newblock \bibinfo{note}{ArXiv:1204.0732 (2012)}.

\bibitem{hofh09}
\bibinfo{author}{Hofheinz, M.} \emph{et~al.}
\newblock \bibinfo{title}{Synthesizing arbitrary quantum states in a
  superconducting resonator}.
\newblock \emph{\bibinfo{journal}{Nature}} \textbf{\bibinfo{volume}{459}},
  \bibinfo{pages}{546--549} (\bibinfo{year}{2009}).

\bibitem{kirc12}
\bibinfo{author}{Kirchmair, G.} \emph{et~al.}
\newblock \bibinfo{title}{Observation of quantum state collapse and revival due
  to the single-photon kerr effect}.
\newblock \bibinfo{note}{ArXiv:1211.2228 (2012)}.

\bibitem{hald99}
\bibinfo{author}{Hald, J.}, \bibinfo{author}{S\o{}rensen, J.~L.},
  \bibinfo{author}{Schori, C.} \& \bibinfo{author}{Polzik, E.~S.}
\newblock \bibinfo{title}{Spin squeezed atoms: A macroscopic entangled ensemble
  created by light}.
\newblock \emph{\bibinfo{journal}{Phys. Rev. Lett.}}
  \textbf{\bibinfo{volume}{83}}, \bibinfo{pages}{1319--1322}
  (\bibinfo{year}{1999}).

\bibitem{geer12}
\bibinfo{author}{Geerlings, K.} \emph{et~al.}
\newblock \bibinfo{title}{Demonstrating a driven reset protocol of a
  superconducting qubit}.
\newblock \bibinfo{note}{ArXiv:1211.0491 (2012)}.

\bibitem{reed10}
\bibinfo{author}{Reed, M.~D.} \emph{et~al.}
\newblock \bibinfo{title}{High-fidelity readout in circuit quantum
  electrodynamics using the jaynes-cummings nonlinearity}.
\newblock \emph{\bibinfo{journal}{Phys. Rev. Lett.}}
  \textbf{\bibinfo{volume}{105}}, \bibinfo{pages}{173601}
  (\bibinfo{year}{2010}).

\end{thebibliography}

%{\bf Supplementary Information} is linked to the online version of the paper at www.nature.com/nature.

%{\bf Author Contributions}

%K.M and S.W. performed the experiment and analyzed the data.  S.W. fabricated the qubit and parametric amplifier. K.M. wrote the manuscript. K.B. helped with the experiment setup, provided theoretical support, and wrote the supplementary information. E.G. conceived of the experiment and provided theoretical support. All work was carried out under the supervision of I.S.

%Reprints and permissions information is available at www.nature.com/reprints

%The authors declare no competing financial interests.

Correspondence and requests for materials should be addressed to katerm@berkeley.edu

%\newpage

%\hspace{-.5in}\includegraphics[angle = 0, width = 1.1\textwidth]{supplementaryinformation_prl}
%\includegraphics{supplementaryinformation_prl.pdf}

%\begin{eqnarray}
%\langle \dot{\sigma}_x \rangle  = -\gamma (N+1/2) \langle \sigma_x\rangle \nonumber\\
%\langle \dot{\sigma}_y \rangle  = -\gamma (N+1/2) \langle \sigma_y\rangle ,\label{eq:obe}\\
%\langle \dot{\sigma}_z \rangle  = -\gamma (2 N+1) \langle \sigma_z\rangle + \gamma  \nonumber.
%\end{eqnarray}

%\begin{eqnarray}
%T_2\rightarrow \frac{T_1}{N-M+1/2}
%\end{eqnarray}

%\begin{eqnarray}
%\langle \sigma_z\rangle\rightarrow \frac{1}{2N+1}
%\end{eqnarray}

%\begin{eqnarray}
%P_e = \frac{N}{2N+1}
%\end{eqnarray}

\end{document}

\title{Supplementary information for  \\ "Suppression of radiative decay of atomic coherence in squeezed vacuum"}

\author{K. W. Murch}
\affiliation{Quantum Nanoelectronics Laboratory, Department of Physics, University of California, Berkeley CA 94720}
\author{S. J. Weber}
\affiliation{Quantum Nanoelectronics Laboratory, Department of Physics, University of California, Berkeley CA 94720}
\author{K. M. Beck}
\affiliation{Department of Physics, MIT-Harvard Center for Ultracold Atoms, and Research Laboratory of Electronics, Massachusetts Institute of Technology, Cambridge, MA 02139}
\author{E. Ginossar}
\affiliation{Advanced Technology Institute and Department of Physics, University of Surrey, Guildford, GU2 7XH, United Kingdom}
\author{I. Siddiqi}
\affiliation{Quantum Nanoelectronics Laboratory, Department of Physics, University of California, Berkeley CA 94720}

\date{\today}
\maketitle

\section{The master equation for the transmon-cavity system}

In the main text, we state that the irreversible dynamics of the effective two-level system (TLS) defined by the ground state and lower polariton of the transmon-cavity system are the same as for an atom interacting with a broadband squeezed vacuum. In this section, we give the theoretical reasoning behind this statement.  We derive the effective master equation for the transmon-cavity system and compare it to previous predictions for an atomic system interacting directly with the squeezed vacuum. This problem was treated before in the  ``bad cavity'' limit where the cavity decay was taken to be the fastest process in the system \cite{Rice_CQED_SQ_1992}. Here, this assumption does not hold and we follow a different derivation.

The system is a transmon circuit strongly coupled to a superconducting cavity. The cavity is also coupled to the line modes in which the squeezed reservoir is realized. The chosen system parameters allow for several theoretical simplifications.
The squeezing bandwidth ($BW=13~\textrm{MHz}$) was 50 times the radiative qubit linewidth ($\gamma/2\pi=$~240~kHz). From the point of view of the qubit, the squeezing is well approximated as having infinite bandwidth. However, since the polariton transitions are well separated with respect to the squeezing bandwidth, with 255~MHz between the $\ket{g}\rightarrow\ket{-}$ and $\ket{g}\rightarrow\ket{+}$ transitions, the squeezed vacuum only addresses one pair of levels, an approximation that we make in Section 2. The several-GHz qubit transition frequencies allow us to use the rotating wave approximation.  We additionally assume the Born approximation, which terminates perturbation theory at the second order, and the Markov approximation, which assumes that the squeezed reservoir has no memory (its correlation time approaches zero).

The Hamiltonian for this system is
\begin{equation}
\label{bare_hamil}
\hamil=\hamil^{TC}+\sum_{k}\omega_{k}b_{k}\hc b_{k}+\sum_{k}\text{h}_{k}\left(b_{k}\hc +b_{k} \right)\left(a\hc+a\right)
\end{equation}
where $\hamil^{TC}$ is the transmon-cavity Hamiltonian, $\omega_{k}$ is the frequency corresponding to the $k^{\text{th}}$ traveling wave line mode, $b_{k}\hc $ ($b_{k}$) is the creation (annihilation) operator for the $k^{\text{th}}$ line mode and $a\hc $ ($a$) is the creation (annihilation) operator for the cavity photon. $\text{h}_{k}$ are the real matrix elements of the transitions between the $k^{\text{th}}$ line mode and the cavity mode. The transmon-cavity Hamiltonian $\hamil^{TC}$ is a generalized Jaynes-Cummings Hamiltonian for the transmon levels and the cavity mode \cite{Koch2007}.

To express the system evolution in terms of the line mode correlations, we first diagonalize the evolution to represent it in a joint polariton and line mode basis \cite{Beaudoin_DRESSED_DEPH_2011}. We apply the transformation $U$ that diagonalizes the closed system part of the Hamiltonian $\hamil^{TC}$  to $\hamil$ %$U$ does not generally have an analytic form \cite{Koch2007}.
\begin{eqnarray}
\hamil' & = & \hamil_{0}+\sum_{k}\text{h}_{k}\left(b_{k}\hc +b_{k}\right)U\left(a\hc + a\right)U\hc \\
 & = & \hamil_{0}+\sum_{k}\text{h}_{k}\left(b_{k}\hc +b_{k}\right)\sum_{i,j>i} \left(A_{ij}\ket{i}\bra{j}+A_{ij}^*\ket{j}\bra{i}\right).
\end{eqnarray}
Here, the free evolution is $\hamil_{0}=\sum_{i}\epsilon_{i}^{TC}\ket{i}\bra{i}+\sum_{k}\omega_{k}b_{k}\hc b_{k}$ and we express the transformed cavity operators in terms of ladder operators between the polariton levels $\ket{i},\ket{j}$ in the diagonalized system $U\left(a+a\hc\right)U\hc=\tilde{a}\hc + \tilde{a}=\sum_{i,j>i}\left(A_{ij}\ket{i}\bra{j}+A_{ij}^*\ket{j}\bra{i}\right)$. This expansion relies the hermiticity of $\tilde{a}\hc+\tilde{a}$ to allow the restriction $j>i$ and since $\tilde{a}\hc + \tilde{a}$ changes the parity of the state, $A_{ii}=0$ \cite{Beaudoin_DRESSED_DEPH_2011}. The energy of the $i^{\text{th}}$ polariton is denoted $\epsilon_{i}^{TC}$.

Now, we move to the interaction picture with respect to $\hamil_0$ where we can apply perturbation theory. The coupling term $\coup=\sum_{i,j>i}\sum_{k}\text{h}_{k}\left(b_{k}\hc +b_{k}\right)\left(A_{ij}\ket{i}\bra{j}+A_{ij}^{*}\ket{j}\bra{i}\right)$ is written as
\begin{eqnarray}
\ipict{\coup}_t & = & e^{\ii\hamil_0 t}\coup e^{-\ii\hamil_0 t}=B_t \hc \sigma_t+B_t \sigma_t \hc.
\end{eqnarray}
 Here, we define a modified line mode photon annihilation operator $B_t=\sum_{k}h_{k}e^{-\ii\omega_{k}t}b_{k}$ and the polarition jump operator $\sigma_t=\sum_{i,j>i}A_{ij}e^{\ii\Delta_{ij}t}\ket{i}\bra{j}$. We have used the rotating wave approximation to remove the terms $B_t\hc \sigma_t\hc$ and $B_t \sigma_t$. The energy difference between polarition levels is written as $\Delta_{ij}=\epsilon_{i}^{TC}-\epsilon_{j}^{TC}$.

In perturbation theory, the system evolution is $\ipict{\dot{\rho}_t}=\ii\left[\ipict{\coup}_t,\ipict{\rho}_t\right]$. Applying this twice, we obtain
\begin{eqnarray}
\ipict{\dot{\rho}}_t & = & \ii\left[\ipict{\coup}_t,\int_{0}^{t}d\tau\ii\left[\ipict{\coup}_{\tau},\ipict{\rho}_{\tau}\right]\right] =  \int_{0}^{t}d\tau\left(\ipict{\coup}_{\tau}\ipict{\rho}_{\tau}\ipict{\coup}_{t}-\ipict{\coup}_{t}\ipict{\coup}_{\tau}\ipict{\rho}_{\tau}+\text{h.c.}\right)
\end{eqnarray}

Taking the partial trace over the line modes, we get the evolution
of $\ipict{\rho}_{t,\tc}=\tr{B}{\ipict{\rho}_{t}}$. To evaluate this,
we expand $\ipict{\coup}$ using Equation~(4) and use the fact that the trace
is cyclic. We further assume the Born approximation \cite{open_quantum_systems} $\ipict{\rho}=\ipict{\rho}_{\tc}\otimes\ipict{\rho}_{B}$ to evaluate
terms like $\tr{B}{B_{\tau}\hc \sigma_{\tau}\ipict{\rho}_{\tau}B_{t}\hc \sigma_{t}}=\sigma_{\tau}\ipict{\rho}_{\tau,\tc}\sigma_{t}\tr{B}{B_{\tau}\hc \ipict{\rho}_{\tau,B}B_{t}\hc}=\sigma_{\tau}\ipict{\rho}_{\tau,\tc}\sigma_{t}\left\langle B_{t}\hc B_{\tau}\hc \right\rangle $.
\begin{eqnarray}
\ipict{\dot{\rho}}_{t,\tc} & = & \int_{0}^{t}d\tau\left\{ \left(\sigma_{\tau}\ipict{\rho}_{\tau,\tc}\sigma_{t}-\sigma_{t}\sigma_{\tau}\ipict{\rho}_{\tau,\tc}\right)\left\langle B_{t}\hc B_{\tau}\hc \right\rangle +\left(\sigma_{\tau}\hc \ipict{\rho}_{\tau,\tc}\sigma_{t}\hc -\sigma_{t}\hc \sigma_{\tau}\hc \ipict{\rho}_{\tau,\tc}\right)\left\langle B_{t}B_{\tau}\right\rangle \right.\\
 &  & \left.+\left(\sigma_{\tau}\hc \ipict{\rho}_{\tau,\tc}\sigma_{t}-\sigma_{t}\sigma_{\tau}\hc \ipict{\rho}_{\tau,\tc}\right)\left\langle B_{t}\hc B_{\tau}\right\rangle +\left(\sigma_{\tau}\ipict{\rho}_{\tau,\tc}\sigma_{t}\hc -\sigma_{t}\hc \sigma_{\tau}\ipict{\rho}_{\tau,\tc}\right)\left\langle B_{t}B_{\tau}\hc \right\rangle +\text{h.c.}\right\}. \nonumber
\end{eqnarray}

Consider one of these terms, $\int_{0}^{t}d\tau\,\sigma_{t}\hc \sigma_{\tau}\ipict{\rho}_{\tau,\tc}\left\langle B_{t}B_{\tau}\hc \right\rangle $.
Expanding the reservoir correlator $\left\langle B_{t}B_{\tau}\hc \right\rangle $:
\begin{eqnarray}
\left\langle B_{t}B_{\tau}\hc \right\rangle  & = & \left\langle \sum_{k}\text{h}_{k}e^{-\ii\omega_{k}t}b_{k}\sum_{k'}\text{h}_{k'}e^{\ii\omega_{k'}\tau}b_{k'}\hc \right\rangle \\
 & = & \sum_{k,k'}\text{h}_{k}\text{h}_{k'}e^{-\ii\left(\omega_{k}t-\omega_{k'}\tau\right)}\tr{B}{\left(b_{k}\hc b_{k'}+\delta_{kk'}\right)\ipict{\rho}_{\tau,B}}\\
 & = & \sum_{k}\left(N_{k}+1\right)\text{h}_{k}^{2}e^{-\ii\omega_{k}\left(t-\tau\right)}
\end{eqnarray}
where $\delta_{kk'}$ is the Kronecker delta. Incorporating this into the full term
\begin{eqnarray}\label{one_term_BBD}
\int_{0}^{t}d\tau\,\sigma_{t}\hc \sigma_{\tau}\ipict{\rho}_{\tau,\tc}\left\langle B_{t}B_{\tau}\hc \right\rangle  & = & \int_{0}^{t}d\tau\,\sum_{i,j>i}A_{ij}^{*}e^{-\ii\Delta_{ij}t}\ket{j}\bra{i} \label{BBD_term}\\
 &  & \times\sum_{i',j'>i'}A_{i'j'}e^{\ii\Delta_{i'j'}\tau}\ket{i'}\bra{j'} \ipict{\rho}_{\tau,\tc}\sum_{k}\left(N_{k}+1\right)\text{h}_{k}^{2}e^{-\ii\omega_{k}\left(t-\tau\right)}\nonumber \\
 & = & \int_{0}^{t}d\tau\,\sum_{i,j>i}\left|A_{ij}\right|^{2}\ket{j}\bra{i}i\rangle\bra{j}\ipict{\rho}_{\tau,\tc}\sum_{k}\left(N_{k}+1\right)\text{h}_{k}^{2}e^{-\ii\left(\omega_{k}+\Delta_{ij}\right)\left(t-\tau\right)} \label{term-2}
\end{eqnarray}
where $N_k=\langle b_k\hc  b_k\rangle$. We will neglect fast-oscillating terms which have little effect on the decay rates by effectively taking $j= j'$. Other terms will give rise to the squeezing correlation function $\langle b_k b_{k'}\rangle=M_k\delta_{\omega_k+\omega_{k'},2\omega_0}$ \footnote{The photons in modes that obey $\omega_k+\omega_{k'}=2\omega_0$ are correlated by the parametric process.}. The sum in this term (\ref{term-2}) is an expression that refers to the line
modes. The temporal kernel in the integral (\ref{one_term_BBD}) can be written in frequency space as
\begin{eqnarray}
K(t-\tau)=\sum_{k}\left(N_k+1\right)\text{h}_{k}^{2}e^{-\ii\left(\omega_{k}+\Delta_{ij}\right)\left(t-\tau\right)} & = & \int_{-\infty}^{\infty}d\omega_{k}\, \DOS{\omega_k}\left(N_{\omega_k}+1\right)\text{h}_{\omega_{k}}^{2}e^{-\ii\left(\omega_{k}+\Delta_{ij}\right)\left(t-\tau\right)}
\end{eqnarray}
where $\DOS{\cdot}$ is the density of states at the given frequency. Since we are interested in the case of a broadband reservoirs $N_{\omega_k}$ has a bandwidth large compared to the typical decay rates and therefore the function $K(t-\tau)$ decays on a timescale of the order of $1/BW$. In the limit of infinite bandwidth it acts similarly to a Dirac delta function effectively restricting the integral to $\tau\approx t$.  This behavior justifies applying the usual set of Born-Markov approximations \footnote{We are neglecting the contributions from the principal part of the temporal integral. These contributions are small for our parameters. %They give rise to frequency shifts of the atomic transitions. % These are very small in our case because of the symmetry of the spectrum.
}, under which the term (\ref{BBD_term}) becomes
\begin{equation}
\sum_{i,j>i}\gamma_{ij}\left|A_{ij}\right|^{2}\sraise \slower \ipict{\rho}_{t,\tc}\left(N_{\Delta_{ij}}+1\right)
\end{equation}
where $\sraise =\ket{j}\bra{i}$ and $\slower =\ket{i}\bra{j}$
with $j>i$ and where $\gamma_{ij}=2\pi\DOS{\Delta_{ij}}h^2(\Delta_{ij})$.

 Computing the other terms similarly and transforming back from the
interaction picture,
%i.e.
%\begin{eqnarray}
%\dot{\rho}_{t,\tc} & = & \frac{d}{dt}\left(e^{-\ii\hamil_{0}t}\ipict{\rho}_{t,\tc}e^{\ii\hamil_{0}t}\right)\\
% & = & -\ii\left[\hamil_{0},\rho_{t,\tc}\right]+e^{-\ii\hamil_{0}t}\ipict{\dot{\rho}}_{t,\tc}e^{\ii\hamil_{0}t}
%\end{eqnarray}
we find
\begin{eqnarray}\label{ME_Transmon_Cavity}
\dot{\rho}_{t,\tc}=\dot{\rho}_{t} & = & -\ii\left[\hamil_{0},\rho_{t}\right]\\
 &  & +\sum_{i,j>i}\gamma_{ij}\left\{ \left(2\sraise \rho \sraise -\sraise \sraise \rho-\rho \sraise  \sraise \right)A_{ij}^{*2}M_{\Delta_{ij}}e^{-2\ii\left(\omega_{0}+\Delta_{ij}\right)t}\right.\nonumber \\
 &  & \left.+\left(2\slower \rho \slower -\slower \slower \rho-\rho \slower \slower \right)A_{ij}^{2}M_{\Delta_{ij}}^{*}e^{2\ii\left(\omega_{0}+\Delta_{ij}\right)t}\right.\nonumber \\
 &  & \left.+\left(2\slower \rho \sraise -\sraise \slower \rho-\rho \sraise \slower \right)|A_{ij}|^{2}\left(N_{\Delta_{ij}}+1\right)\right.\nonumber \\
 &  & \left.+\left(2\sraise \rho \slower -\slower \sraise \rho-\rho \slower \sraise \right)|A_{ij}|^{2} N_{\Delta_{ij}}\right\} \nonumber
\end{eqnarray}
where $\omega_0$ is the center frequency of the squeezed reservoir. This equation is the multi-level generalization of the two-level master equation derived by Gardiner, \texteq{10} of \textref{\cite{Gardiner86}} \footnote{The sign difference between Gardiner's master equation and \texteq{\ref{ME_Transmon_Cavity}} in the $M$-dependent terms comes from writing the bare Hamiltonian \texteq{\ref{bare_hamil}} with $a+ a\hc$ instead of $\ii \left(a-a\hc\right)$}.

\section{Detuning dependence of polariton coherence decay}

When we restrict the master equation (\ref{ME_Transmon_Cavity}) to only two levels, we obtain a mathematically equivalent master equation to \textref{\cite{Gardiner86}}.  Physically, the two lowest levels $i=0,1$ refer to the ground state $|g\rangle$ and the dressed state $|-\rangle$, respectively.  In the experiment, the squeezed vacuum bandwidth is centered around the $|g\rangle\rightarrow |-\rangle$  transition frequency. Since all other transitions are well outside this bandwidth, we can limit ourselves to the two-level dissipative dynamics and directly apply Gardiner's theory to the ground state and lower polariton of the circuit QED system explored experimentally in the main text.

In the absence of a classical coherent drive, the decay of atomic polarizations obey on average

\begin{eqnarray}
&&\dot{\sigma}_{+}=-\Gamma_N \sigma_{+}+\Gamma_M e^{2\ii\delta t} \sigma_{-} \label{Atom-DecayP} \\
&&\dot{\sigma}_{-}=-\Gamma_N \sigma_{-}+\Gamma_M^* e^{-2\ii\delta t}\sigma_{+} \label{Atom-DecayM}
\end{eqnarray}
where $\sigma_{\pm}=\langle S^{\pm}_{01}\rangle$, $\Gamma_N=\gamma(N+1/2)+\gamma_\phi$, $\Gamma_M=\gamma M$, $\gamma=|A_{01}|^2\gamma_{01}$,  and $\delta=\omega_0+\Delta_{01} = \omega_0+\omega_\mathrm{q}$. $\gamma_\phi$ is a phenomenological pure dephasing that was measured independently.

The decay dynamics of the atomic coherences $\sigma_{x,y}$ can be found by solving Equations (\ref{Atom-DecayP}) and (\ref{Atom-DecayM}) and then transforming to a frame rotating with angular frequency $\delta$, i.e. $\sigma_{\pm}(t)=e^{\pm i\delta t}\tilde{\sigma}_{\pm}(t)$. These dynamics will have two decay timescales $\Gamma_N\pm \sqrt{|\Gamma_M|^2-\delta^2} = \gamma(N+1/2)+\gamma_\phi \pm\sqrt{\gamma^2 |M|^2-\delta^2}$ unless $\delta =0$ or $|\delta|>\gamma|M|$.  Detuning also leads to an effective time-dependent rotation of the squeezing ellipse. This limits the influence of squeezing on decay to the range $|\delta| \le \gamma|M|$ even in the ideal case of infinite bandwidth squeezing.

\begin{figure*}
 \begin{center}
\includegraphics[width=5.5in]{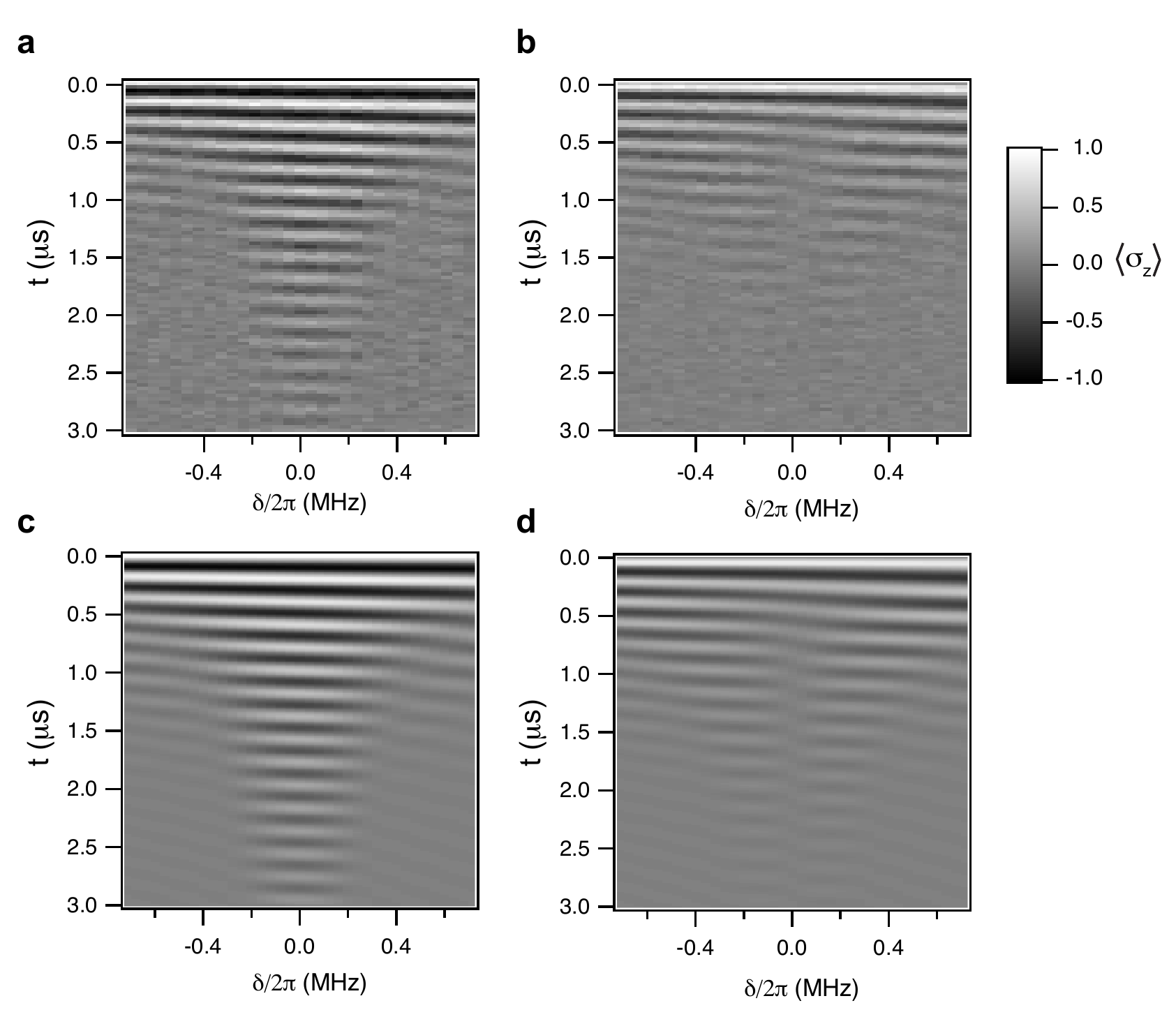}
\caption{Comparison of measured and calculated decay of atomic coherence versus detuning.  {\bf a,b}  Ramsey measurements as a function of detuning, $\delta = \omega_0 - \omega_\mathrm{q}$ for the qubit prepared initially along $|\pi/2,\pi/2\rangle$ and $|\pi/2,0\rangle$ respectively. {\bf c,d} Corresponding simulated Ramsey measurements for the qubit prepared initially along $|\pi/2,\pi/2\rangle$ and $|\pi/2,0\rangle$ respectively. \label{fig:s1}}
\end{center}
\end{figure*}

To interpolate between the limits of large detuning ($|\delta|>\gamma|M|$) and $\delta=0$ where the decay is described by a single exponential time constant, Figure 4a in the main text presents the effective decay constants that were obtained by fitting the Ramsey measurements to a single exponential decay at each detuning. For comparison to the theoretical prediction, the same fitting routine was applied to theoretical decay curves given by Equations (\ref{Atom-DecayP}) and (\ref{Atom-DecayM}) by transforming into the frame rotating with angular frequency $\delta+\omega_\mathrm{mod}$, with $\omega_\mathrm{mod}/2\pi = 5$ MHz and $\gamma_\phi = 2\pi\times 2.4\times10^{-4}$ s$^{-1}$. The decay constants that result from this fit are plotted as the theoretical curve in Figure 4a. For a direct comparison to theory, we present experimental data and the predicted Ramsey measurement results versus detuning in Figure \ref{fig:s1}.

